\documentstyle[subfigure,preprint,aps,epsfig]{revtex}

\begin{document}

\preprint{\begin{tabular}{l}
\hbox to\hsize{  \hfill KAIST--TH 2001/05}\\[-3mm]
\hbox to\hsize{  \hfill DESY 01--033}\\[5mm] \end{tabular} }

\title{Radion Phenomenology in the \\ Randall Sundrum Scenario}

\author{S. Bae, P. Ko, H. S. Lee}

\address{
Department of Physics, KAIST, Taejon, 305-701, Korea
\\E-mail: pko@muon.kaist.ac.kr
}

\author{Jungil Lee}

\address{
Deutsches Elektronen-Synchrotron DESY, D-22603 Hamburg, Germany
\\E-mail: jungil@mail.desy.de
}


\maketitle

\begin{abstract}
Phenomenology of a radion ($\phi$) in the Randall-Sundrum scenario is 
discussed. The radion couples to the trace of energy momentum tensor of 
the standard model with a strength suppressed only by a new scale 
($\Lambda_{\phi}$) which is an order of the electroweak scale. 
In particular, the effective coupling of a radion to two gluons is 
enhanced due to the trace anomaly of QCD. 
Therefore,  its production cross section  at hadron colliders could be 
enhanced, and the dominant decay mode of a relatively light radion is  
$\phi \rightarrow g g$, unlike the SM Higgs boson case. 
We also present constraints on the mass $m_{\phi}$ and the new scale 
$\Lambda_{\phi}$ from the Higgs search limit at LEP, perturbative 
unitarity bound and the stability of the radion/Higgs potential under
radiative corrections. 
\end{abstract}

\section{Introduction}
 
Despite the tremendous success, the standard model (SM) has several 
theoretical drawbacks, one of which is related with stabilizing the
electroweak scale relative to the Planck scale under quantum corrections, 
which is known as the gauge hierarchy problem. 
Traditionally, there have been basically two avenues to solve this problem : 
(i) electroweak gauge symmetry is spontaneously broken by some new strong 
interactions (technicolor or its relatives) or (ii) there is a supersymmetry  
(SUSY) which is spontaneously broken in a hidden sector, and superpartners 
of SM particles have masses around the electroweak scale $O(100-1000)$ GeV. 
However, new mechanisms based on the developments in superstring and M 
theories including D-branes have been suggested by Randall and Sundrum 
\cite{randsun}. If our world is confined to a three-dimensional brane
and the warp factor in the Randall and Sundrum (RS) theory is much smaller
than 1, then loop corrections cannot destroy the mass hierarchy 
derived from  the relation $v=e^{-k r_c \pi} v_0$,  where $v_0$ is the VEV 
of Higgs field  ($\sim$ $O(M_P)$) in the 5-dimensional RS theory, 
$e^{-k r_c \pi}$ is the warp factor, and $v$ is the VEV of Higgs field 
($\sim$ 246 GeV) in the 4-dimensional effective theory of the RS theory by 
a kind of dimensional reduction.  Especially the extra-dimensional subspace 
need not be a circle $S^1$ like the Kaluza-Klein theory \cite{randsun}, 
and in that case, it is crucial to have a mechanism to stabilize the modulus.  
One such a mechanism was recently proposed by Goldberger and Wise 
\cite{wise,wise2}, and also by Cs${\acute{\rm a}}$ki {\it et al}.
\cite{csaki}. 
In such a case, the modulus (or the radion $\phi$ from now on) is likely to 
be lighter than the lowest Kaluza-Klein excitations of bulk fields. Also its 
couplings to the SM fields are completely determined by general 
covariance in the four-dimensional spacetime, as shown in Eq.~(1) below. 
If this scenario is realized in nature, this radion could be the first 
signature of this scenario, and it would be important to determine its 
phenomenology at the current/future colliders, which is the purpose of this 
talk~\cite{pko}. Some related issues 
were addressed in Refs.~\cite{mahanta,wells}.

In this talk, we first recapitulate the interaction Lagrangian for 
a single radion and the SM fields in the Randall Sundrum scenario, and 
discuss the decay rates and the branching ratios of the radion into SM 
particles.  Then the perturbative unitarity bounds and stability condition
for the radion Higgs potential on the radion mass $m_{\phi}$ and 
$\Lambda_{\phi}$ are considered.  Current bounds on the SM Higgs search 
can be easily translated into the corresponding bounds on the radion.
Then the radion production cross sections at next linear colliders (NLC's) 
and hadron colliders such as the Tevatron and LHC are considered. 
Then our results will be summarized at the end.

\section{Radion in the Randall Sundrum Scenario I}

In the RS theory I \cite{randsun}, the hierarchy problem between 
the Planck scale $\Lambda_{\rm Pl} \sim 10^{18}$ GeV and the electroweak scale
$\Lambda_{\rm EW}\sim 10^2$ GeV can be solved geometrically via $v=e^{-kr_c
\pi} v_0$ with $kr_c \sim 12$, where the warp factor $e^{-kr_c \pi}$ is
in the classical RS metric, 
\begin{equation}
ds_{\rm RS}^2=e^{-2kr_c|y|} \eta_{\mu \nu}dx^{\mu}dx^{\nu} -r_c^2dy^2. 
\end{equation}
The radius $r_c$ is the VEV of the $yy$-component of metric $G_{ab}(x, y)$. 
When the bulk field $G_{yy} (x,y)$ is decomposed into Kaluza-Klein (KK) modes,
the lowest lying mode is the massless radion ($\phi$) in
the original RS model I, since there is  no tree-level 
stabilization mechanism for the radion. The loop corrections can solve the 
hierarchy problem, but give too light radion to be consistent with 
experiments \cite{garriga}. Therefore, a tree-level stabilization mechanism 
is needed, and the Goldberger-Wise mechanism \cite{wise} is a promising one, 
because they stabilized the modulus without any severe fine-tuning of the
parameters in the full theory. 

In the Goldberger-Wise stabilization mechanism,
there is a bulk scalar field $\Phi(x, y)$ which has large quartic
self-interactions only on the hidden and visible branes, and an
extra-dimension dependent VEV ${\tilde \Phi}(y)$. After replacing the field
$\Phi(x, y)$ in the original Lagrangian with its VEV ${\tilde \Phi}(y)$
and integrating the Lagrangian over $y$, we have the modulus stabilizing
potential. Since there is a potential stabilizing the radion, the radion has a
mass of order $O(10)$ GeV at the tree level \cite{wise2}. From a recent
analysis \cite{bae}, it was known that the one-loop allowed radion mass can 
be $\sim O(10)$ times larger than the tree-level one, but the radion is still 
lighter than the Kaluza-Klein modes.

\section{Scale Anomaly and the Interaction Lagrangian for a Radion}
\subsection{Scale Anomaly}

If a Lagrangian possesses no dimensionful parameter, there is a scale symmetry
at classical level, for which one can construct a conserved Noether current 
$D_{\mu}$. One can show that the improved energy momentum tensor
$\Theta_{\mu\nu}$ satisfies the following relation :
\begin{equation}
\partial_{\mu} D^{\mu} = \Theta_{\mu}^{\mu}.
\end{equation}
For example, if one considers QCD as an example, the improved energy 
momentum tensor $\Theta^{\mu\nu}$ is given by 
\begin{eqnarray}
\Theta^{\mu\nu} & = &  -F^{a \lambda \mu} F^{a\nu}_\lambda+ \frac{1}{4}g^{\mu
\nu} F^{a \rho \sigma}F^a_{\rho \sigma}  
\\
& + & \frac{i}{2} 
\bar{q}(\gamma^\mu D^\nu + \gamma^\nu D^\mu) q 
- g^{\mu\nu} \bar{q} \left( i \gamma^{\alpha} D_{\alpha} - m_q \right) q.
\nonumber 
\end{eqnarray}
It is clear that this theory has traceless energy momentum tensor at 
classical level if we consider the massless quark limit (without dimensionful
parameters).

However, this is no longer true in quantum field theory (QFT) in which the
radiative corrections are usually divergent so that renormalization procedure
is called for. In the renormalization, one needs to regularize the theory
with a suitable cutoff parameter to make loop integrals finite. Therefore
a hidden cutoff scale is necessarily involved with QFT.  If we use the
dimensional regularization instead of cutoff regularization, the dimensionless
parameter in four-dimensional theory is no longer dimensionless in 
$D$-dimensional theory. And the classical scale symmetry in $4-D$ is no 
longer  a good symmetry in arbitrary $D$ dimensions.

To look into the energy momentum tensor $\Theta_{\mu\nu}$  more closely at
quantum level, let us consider the scale dependence of quantum effective 
action. Since the renormalized coupling $g_s (\mu)$ has a scale dependence,
\begin{equation}
\mu {{\partial g_s} \over {\partial \mu}}  =  \beta ( g_s ),
\end{equation}
one  finds that
\begin{equation}
\mu {\partial {\cal L} \over \partial \mu} = \beta ( g_s ) 
{\partial {\cal L} \over \partial g_s} = \Theta^\mu_\mu .
\end{equation}
Therefore we end up with
\begin{equation}
\Theta_\mu^\mu ({\rm SM})^{\rm anom} = \sum_{G={\rm SU}(3)_C, \cdots } 
{\beta_G (g_G ) \over 2 g_G}~
{\rm tr} (F_{\mu\nu}^G F^{G \mu\nu}).
\label{eq:anom} 
\end{equation}
Since the classical scale symmetry is broken by quantum effects, it is called
scale anomaly (like the chiral anomaly) \cite{trace}. 
If there were mass parameters in the classical Lagrangian, the scale symmetry
would have been broken already at classical level so that 
$\Theta_{\mu}^{\mu}$ is not zero and this should be considered altogether with
the scale anomaly term. 

\subsection{Interaction Lagrangian for a Radion}

The interaction of the radion with the SM fields at an electroweak scale is 
dictated by the 4-dimensional general covariance, and  is described by the 
following effective Lagrangian via canonical normalizations of the fields
\cite{csaki,wise2} :
\begin{equation}
{\cal L}_{\rm int} =  {\phi \over \Lambda_{\phi}}~{\Theta_\mu^\mu} 
({\rm SM}) + ...,
\end{equation} 
where $\Lambda_{\phi} = \langle \phi \rangle \sim O(v)$. The radion becomes 
massive after the modulus stabilization, and its mass $m_{\phi}$ is a free 
parameter of electroweak scale \cite{csaki}. Therefore, two parameters 
$\Lambda_{\phi}$ and $m_{\phi}$ are required in order to discuss productions 
and decays of the radion at various settings. The couplings of the radion 
with the SM fields look like those of the SM Higgs, except for 
$v \rightarrow \Lambda_{\phi}$. 
However, there is one important thing to be noticed : the quantum corrections
to the trace of the energy-momentum tensor lead to trace anomaly, leading to
the additional effective radion couplings to gluons or photons in addition 
to the usual loop contributions. This trace anomaly contributions will lead
to distinct signatures of the radion compared to the SM Higgs boson.

The trace of energy-momentum tensor of the SM fields at tree level is
easily derived :
\begin{eqnarray}
{\Theta_\mu^\mu} ({\rm SM})^{\rm tree} & = & 
\sum_{f} m_f \bar{f} f - 2 m_W^2 W^{+}_{\mu} W^{- \mu} - 
m_Z^2 Z_{\mu} Z^{\mu} 
\nonumber 
\\
& + &
\left( 2 m_h^2 h^2 - \partial_{\mu} h \partial^{\mu} h \right) + ... ,
\end{eqnarray}
where we showed terms with only two SM fields, since we will discuss two 
body decay rates of the radion into the SM particles, except the gauge bosons 
of which virtual states are also considered. The couplings between the radion
$\phi$ and a fermion pair or a weak gauge boson pair are simply related with 
the SM Higgs couplings with these particles through simple rescaling : 
$g_{\phi-f-\bar{f}} = g_{h-f-\bar{f}}^{\rm SM}~{v/\Lambda_{\phi}}$, and so on.
On the other hand, the $\phi-h-h$ coupling is more complicated than the SM
$h-h-h$ coupling. There is a momentum dependent part from the derivatives 
acting on the Higgs field, and this term can grow up as the radion mass gets 
larger or the CM energy gets larger in hadroproductions of the radion. It 
may lead to the violation of perturbative unitarity, which will be addressed
after we discuss the decay rates of the radion. Finally, 
there is no $h-\phi-\phi$ coupling,  
since the radion couples to the trace of the energy momentum tensor and 
there should be no $h-\phi$ mixing after field redefinitions in terms of 
physical fields. 

In addition to the tree-level $T_{\mu}^{\mu} ({\rm SM})^{\rm tree}$, 
there is also the trace anomaly term  for gauge fields \cite{trace}, 
Eq.~(\ref{eq:anom}).  
This trace anomaly couples with the parameter of conformal transformations 
in our 3-brane. And the radion $\phi$ plays the same role as the parameter of 
conformal transformation, since it belongs to the warp factor in the 5 
dimensional RS metric \cite{csaki}. Therefore, the parameter associated with 
the conformal transformation is identified with the radion field $\phi$. As a 
result, the radion $\phi$ has a coupling to the trace anomaly term. For QCD 
sector as an example, one has
\begin{equation}
{\beta_{QCD} \over 2 g_s} = - \left( 11 - {2 \over 3} n_f \right) 
{\alpha_s \over 8 \pi} \equiv - {\alpha_s \over 8 \pi}~b_{QCD},
\end{equation}
where $n_f = 6 $ is the number of active quark flavors. 
There are also counterparts in the $SU(2) \times U(1)$ sector.
This trace anomaly has an important phenomenological consequence.
For a relatively light radion, the dominant decay mode will not be 
$\phi \rightarrow b \bar{b}$ as in the SM Higgs, but $\phi \rightarrow gg$.

\section{Radion Phenomenology}
\subsection{Decay Properties of a Radion}

Using the interaction Lagrangian Eq.~(7), it is straightforward to calculate 
the decay rates and branching ratios of the radion $\phi$ into 
$f \bar{f}, W^+ W^-, Z^0 Z^0, g g$ and $h h$. 
\begin{eqnarray}
\Gamma ( \phi \rightarrow f \bar{f} ) &=& N_c \frac{ m_f^2 m_{\phi}}
 { 8 \pi \Lambda_\phi^2} \left( 1 - x_f \right)^{3/2}, 
\nonumber 
\\
\Gamma( \phi \rightarrow W^+ W^- ) &=& \frac { m_\phi^3}
{16 \pi \Lambda_\phi^2} \sqrt{ 1 - x_W} \ ( 1 - x_W + \frac{3}{4} x_W^2 ), 
\nonumber   \\
 \Gamma( \phi \rightarrow Z Z ) &=& \frac { m_\phi^3}
{32 \pi \Lambda_\phi^2} \sqrt{ 1 - x_Z} \ ( 1 - x_Z + \frac{3}{4} x_Z^2 ),
\nonumber  \\
 \Gamma( \phi \rightarrow h h ) &=& \frac { m_\phi^3}{32 \pi 
\Lambda_\phi^2} \sqrt{ 1-x_h} \  ( 1 + \frac {x_h}{2} )^2,  
\nonumber  \\
\Gamma( \phi \rightarrow g g ) &=& \frac{ \alpha_s^2 m_\phi^3 }
{32 \pi^3 \Lambda_\phi^2} \left| b_{QCD} + \sum_q I_q( x_q ) \right|^2, 
\end{eqnarray}
where $ x_{f,W,Z,h} = 4 m_{f,W,Z,h}^2/m_\phi^2 $, and 
$I ( z ) = z [ 1 + (1 - z ) f(z) ]$ with
\begin{eqnarray} \label{ffnt}
f(z) & = &  - {1\over 2}~\int_0^1 {dy \over y}~\ln [ 1 - {4 \over z} y (1-y)]
\nonumber
\\
& = & \left\{  \begin{array}{cl}
           {\rm arcsin}^2(1/\sqrt{z}) \,,   &\qquad z \geq 1\,, \\
           -\frac{1}{4}\left[\ln \left(\frac{1+\sqrt{1-z}}{
                                             1-\sqrt{1-z}}\right)
                             -i\pi\right]^2\,, & \qquad z\leq 1\,.
               \end{array} 
\right.
\end{eqnarray}
Note that as $m_{t} \rightarrow \infty$, the loop function approaches
$I( x_t ) \rightarrow 2/3$ so that the top-quark effect decouples and one
is left with $b_{QCD}$ with $n_f = 5$.  For $\phi \rightarrow WW,ZZ$, we have 
ignored $SU(2)_L \times U(1)_Y$ anomaly, since these couplings are allowed 
at the  tree-level already, unlike the $\phi-g-g$ or $\phi-\gamma-\gamma$
couplings. This should be a good approximation for a relatively light radion.

\begin{figure}
\centerline{\epsfxsize=7.5cm \epsfbox{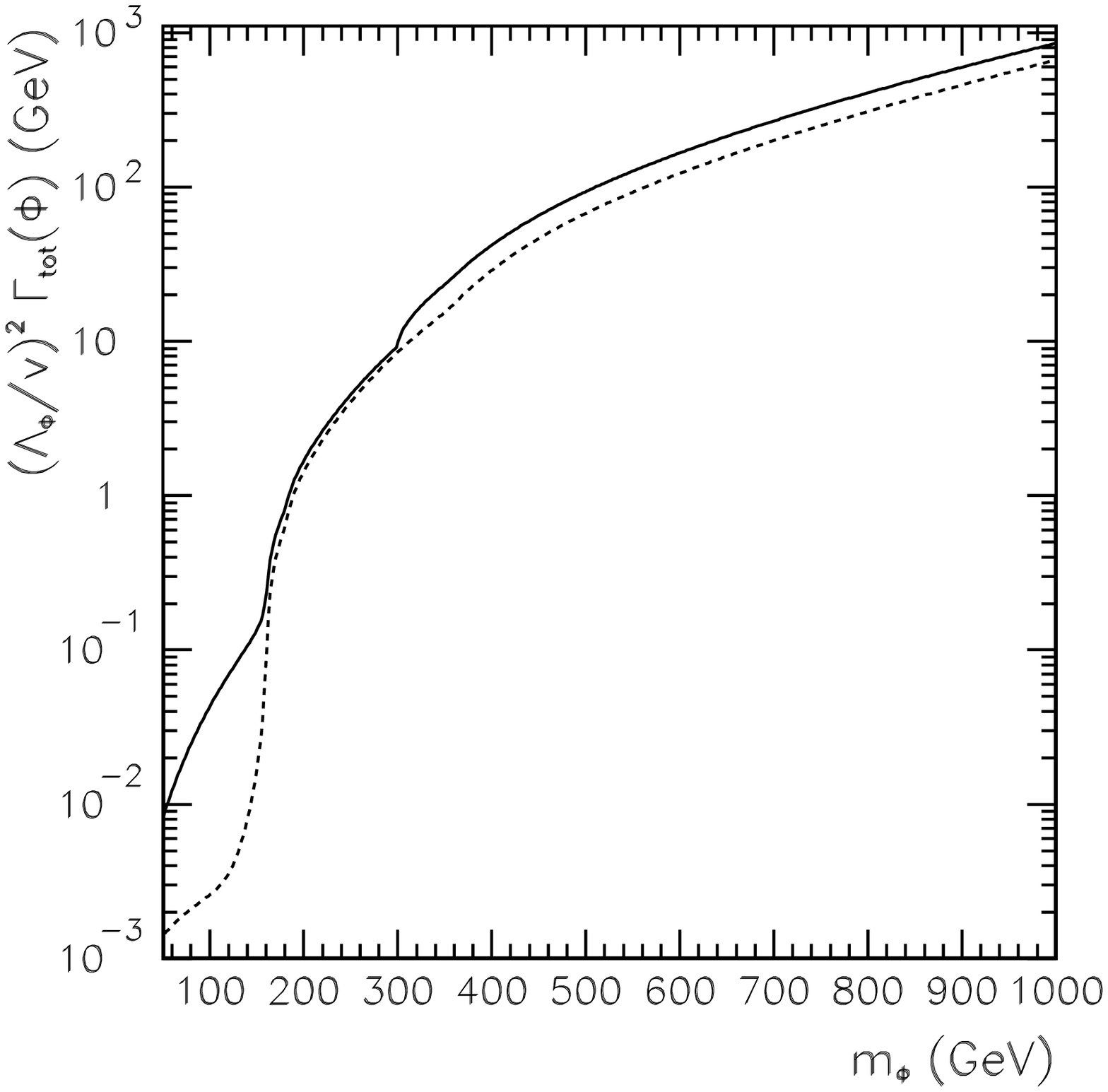}}  
\caption{The total decay rate (in GeV) for the radion $\phi$ for $m_h$=150 GeV
with a scale factor $(\Lambda_{\phi} / v )^2$.
The decay rate of the SM Higgs boson is shown in the dashed curve for 
comparison.
}
\label{decay}

\centerline{\epsfxsize=7.5cm \epsfbox{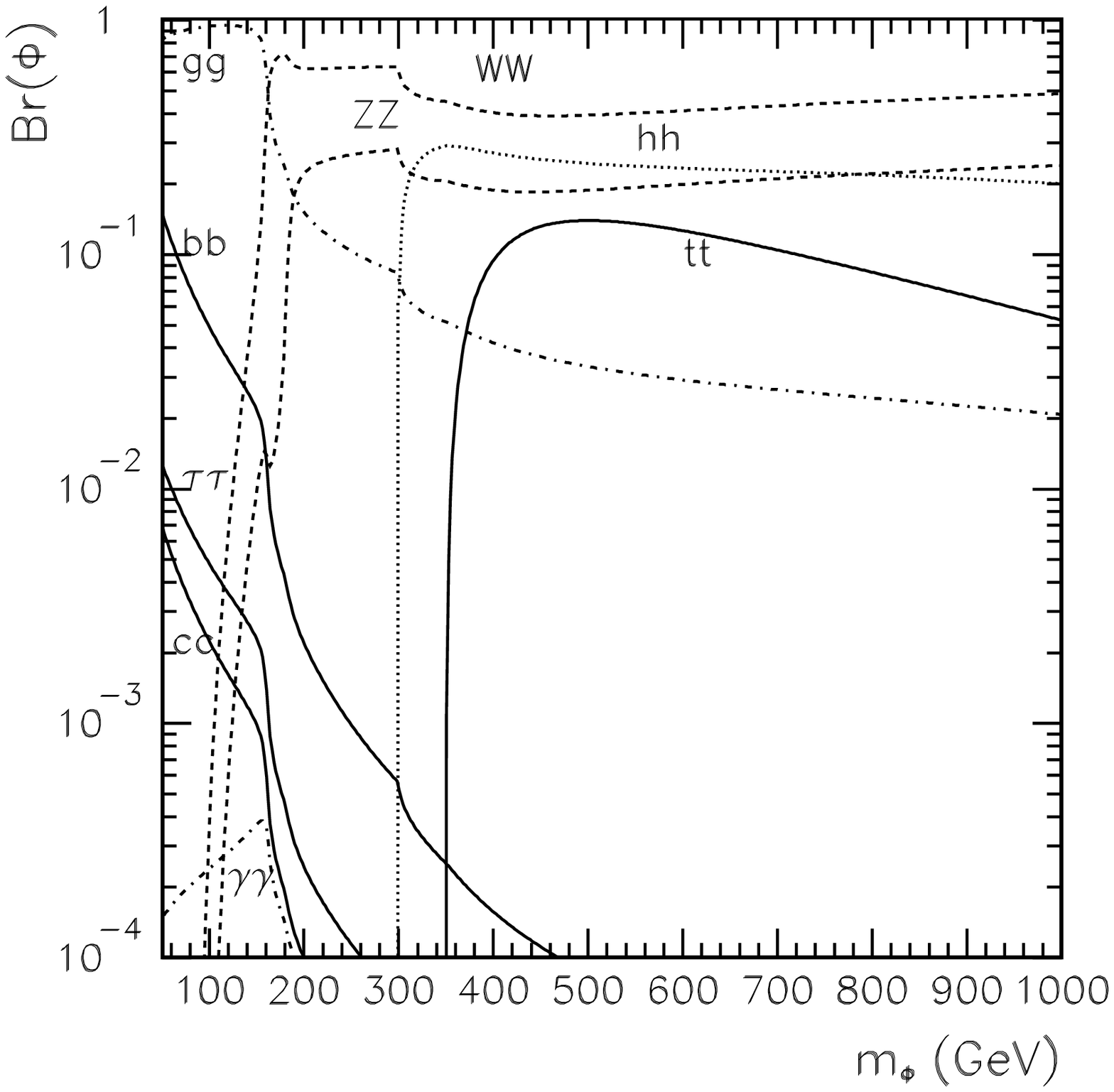}} 
\caption{The branching ratios for the radion $\phi$ into the SM particles.} 
\label{br}
\end{figure}

Using the above results, we show the decay rate of the radion and the 
relevant branching ratio for each channel available for a given $m_{\phi}$
in Figs.~\ref{decay} and \ref{br}.

In the numerical analysis, we use $\Lambda_{\phi} = v = 246$ GeV and 
$m_h = 150$ GeV, and also included the QCD corrections. The decay rates for 
different values of $\Lambda_{\phi}$ can
be obtained through the following  scaling : $\Gamma (\Lambda_{\phi})  
= (v/\Lambda_{\phi})^2 \Gamma (\Lambda_{\phi} = v)$. The decay rate scales 
as $(v/\Lambda_{\phi})^2$, but the branching ratios are independent of 
$\Lambda_{\phi}$.  In Fig.~\ref{decay}, we also show the decay rate of the 
SM Higgs boson with the same mass as $\phi$. We note that the light radion 
with $\Lambda_{\phi} = v$ could be a much broader resonance compared to the 
SM Higgs even if $m_{\phi} \lesssim 2 m_{W}$. This is because the 
dominant decay mode is $\phi \rightarrow g g $ (see Fig.~\ref{br}), unlike 
the SM Higgs for which the $b \bar{b}$ final state is a dominant decay mode. 
This phenomenon is purely a quantum field theoretical effect : enhanced 
$\phi-g-g$ coupling due to the trace anomaly. For a heavier radion, it turns 
out that $\phi \rightarrow V V $ with $V=W$ or $Z$ dominates other decay 
modes once it is kinematically allowed. Also the branching ratio for 
$\phi \rightarrow h h$ can be also appreciable if it is kinematically allowed.
This is one of the places where the difference between the SM and the radion 
comes in.
If $\Lambda_{\phi} \gg  v$, the radion would be a narrow resonance and should
be easily observed as a peak in the two jets or $WW(ZZ)$ final states.
Especially $\phi \rightarrow Z Z \rightarrow (l\bar{l}) (l^{'} \bar{l^{'}})$
will be a gold-plated mode for detecting the radion as in the case of the SM
Higgs. Even in this channel, one can easily distinguish the radion from the 
SM Higgs by difference in their decay width. 

\subsection{Perturbative Unitarity}

\begin{figure}
\centerline{\epsfxsize=7.5cm \epsfbox{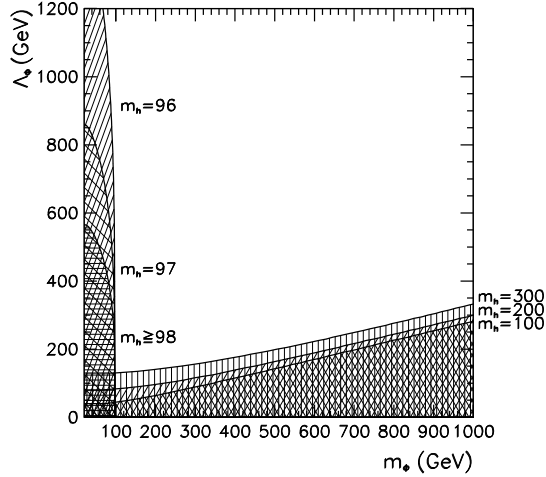}}
\caption{
The excluded region in the $m_\phi$ and $\Lambda_\phi$ space obtained from 
the recent L3 result on the SM Higgs search (the left three curves)  
and perturbative unitarity bound (the lower three curves). }
\label{bound}
\end{figure}

The perturbative unitarity can be violated (as in the SM) in the 
$V_L V_L \rightarrow V_L V_L$ or $h h \rightarrow V_L V_L$, etc. Here
we consider $h h \rightarrow h h$, since the $\phi-h-h$ coupling scales
like $s/\Lambda_{\phi}$ for large $s \equiv ( p_{h_1} + p_{h_2} )^2$.
The tree-level amplitude for this process is 
\begin{eqnarray}
{\cal M} (h h \rightarrow h h) &=& -6\lambda 
-\frac{1}{\Lambda_{\phi}^2} \left(
\frac{s^2}{s-m_{\phi}^2}+\frac{t^2}{t-m_{\phi}^2}+
\frac{u^2}{u-m_{\phi}^2} \right)
\nonumber
\\
&&-36\lambda^2 v^2 \left( \frac{1}{s-m_h^2}+ \frac{1}{t-m_h^2}
+\frac{1}{u-m_h^2} \right)
\end{eqnarray}
where $\lambda$ is the Higgs quartic coupling, and $s+t+u = 4 m_h^2$.
Projecting out the $J=0$ partial wave component ($a_0$) and imposing
the partial wave unitarity condition $| a_0 |^2 \leq {\rm Im} (a_0)$ (i.e.
$| {\rm Re} (a_0)|\leq 1/2$), we get 
the following relation among $m_h, v, m_{\phi}$ and $\Lambda_{\phi}$, for 
$s \gg m_h^2,~ m_{\phi}^2$ :
\begin{equation}
\left|\frac{2m_h^2+m_{\phi}^2}{8\pi\Lambda_{\phi}^2}+\frac{3\lambda}{8\pi}
\right| \leq \frac{1}{2}.
\end{equation}
This bound is shown in the lower three curves of Fig.~\ref{bound}. We note 
that the perturbative unitarity is broken for relatively small 
$\Lambda_{\phi} \lesssim 130 ~(300)$ GeV for $m_{\phi} \sim$ 200 GeV (1 TeV).
Therefore, the tree level results should be taken with care for this range of
$\Lambda_{\phi}$ for a given radion mass. 

\subsection{Stability of the radion and Higgs potential under radiative
corrections}

Adding the loop corrections from the scalar, fermion
vector and KK-mode sectors, the final effective potential up to the one-loop
level \cite{bae} is
\begin{eqnarray}
\label{Veff}
& & V_{\rm eff}(h, \phi)
\nonumber 
\\
& = & 
V_{\rm tree}(h, \phi) +V_{\rm 1 \, loop}^{\rm KK}(\phi) 
+ V_{\rm 1 \, loop}^{\rm scalar}(h, \phi)
+V_{\rm 1 \, loop}^{\rm fermion}(h) + V_{\rm 1 \, loop}^{\rm vector}(h) 
\nonumber
\\
&=&
V_{\phi}(\phi)+\frac{\lambda}{4} \left(h^2 - v_0^2 
e^{- 2 \phi/\Lambda_{\phi}} \right)^2 
+ \delta V_{\rm TeV} e^{-4\phi/\Lambda_{\phi}}~
\nonumber
\\
& + &\frac{1}{4(4\pi)^2} 
\left\{ 
\lambda^2 \left(3h^2 - v_0^2 e^{- 2 \phi/\Lambda_{\phi}} \right)^2
\left( \log \left[ \lambda \left(3h^2 - v_0^2 e^{- 2 \phi/\Lambda_{\phi}} 
\right)/{\mu}^2 \right] - \frac{3}{2} \right) \right\}  
\nonumber
\\
&&
\left.
~~~~+ 3 \lambda^2 \left(h^2 - v_0^2 e^{- 2 \phi/\Lambda_{\phi}} 
\right)^2 \left( \log \left[ \lambda \left(h^2 - v_0^2 e^{- 2 \phi/
\Lambda_{\phi}} \right)/{\mu}^2 \right] - \frac{3}{2} \right) \right\}  
\\
&  + & \frac{1}{(4 \pi)^2} \left\{
-3T^2\left(\log\frac{T}{\mu^2}-\frac{3}{2} \right)
+\frac{3}{2}W^2 \left(\log\frac{W}{\mu^2}-\frac{5}{6} \right)
+\frac{3}{4} Z^2 \left(\log\frac{Z}{\mu^2}-\frac{5}{6} \right)
\right\} .
\nonumber
\end{eqnarray}
The procedure to determine $m_\phi$ and $\Lambda_\phi$ is to  
find a parameter point of ($v_0,~ v_v,~ v_h,~ m,~\delta V_{\rm TeV}$)
which gives a stable vacuum satisfying the two constraints, warp factor
$e^{-kr_c\pi}=O(M_W/M_{P})$ and Higgs VEV $v \simeq 246$ GeV 
and next determine the values of $m_\phi$ and $\Lambda_\phi$ from the 
relations to the previous parameters.

\begin{figure}
\centerline{\epsfxsize=7.5cm \epsfbox{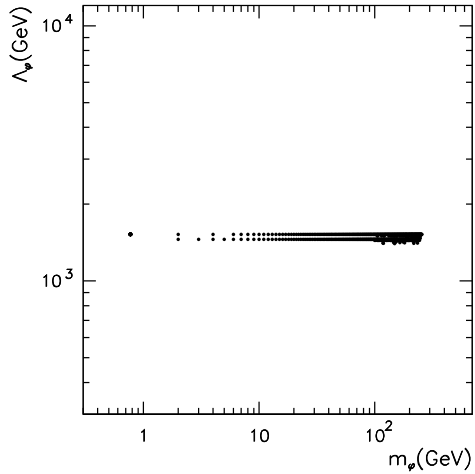}}
\caption{The allowed points for the one-loop effective potential
($M_{(5)}=0.8M_{\rm Pl}$ and $m_h=125$ GeV).}
\label{loop1}
\end{figure}
For the one-loop effective potential, the allowed region for the 5-dimensional
Planck mass $M_{(5)}=0.8M_{\rm Pl}$ and the Higgs mass $m_h=125$ GeV is
$0.8~{\rm GeV} \lesssim m_{\phi} \lesssim 260~{\rm GeV}$ and
$1400~{\rm GeV} \lesssim \Lambda_{\phi} \lesssim 1500~{\rm GeV}$
from Fig.~\ref{loop1} (the central value of $kb_0/2$ is still about 36). 
Almost all the data are focused at
$\Lambda_\phi=1490$ GeV like the tree level case. Therefore, we can conclude
that the {\it naturally} allowed regions are similar for the tree and one-loop
cases, and there is a small shift of the central value of
$\Lambda_\phi$ (or equivalently $kb_0/2$) due to the one-loop
corrections. When $\delta V_{\rm TeV}^{\rm KK}$ and other
parameters are changed continuously, the allowed region in Fig.~\ref{loop1}
can be broader in $\Lambda_\phi$. From numerical analysis, we have found
that only the negative values of $\delta V_{\rm TeV}^{\rm KK}$
can be allowed. Negative values of the
tension shift $\delta V_{\rm TeV}^{\rm KK}$ produce many parameter
points which were not allowed at the tree level, and considerable parts
of these new points make the radion mass sufficiently larger
than the tree-level upper bound of the mass.
It is phenomenologically noteworthy that the one-loop upper bound of the
radion mass $m_\phi$ is rather larger than the tree-level one by
about five times.
But the radion is still the first signal of the RS theory lighter than the
lowest-lying KK mode with a mass of order $O(1)k e^{-kb_0/2} 
\simeq 0.8 \Lambda_\phi$ \cite{{GW7218},{GhPo}}, because the radion mass is
smaller than about $\Lambda_\phi/6$. Due to $m_\phi \lesssim 260$ GeV, 
the branching ratios of the radion into gluon or $W$ boson pairs are
dominant according to the discussions in the previous subsection. 

\subsection{Radion Productions at hadron and linear colliders}

The production cross sections of the radion at hadron colliders are given by
the gluon fusion into the radion through quark-loop diagrams, as in the case
of Higgs boson production, and also through the trace anomaly term, Eq.~(4),
which is not present in the case of the SM Higgs boson :
\begin{equation}
\sigma ( p p \, ({\rm or}~ p \bar{p}) \rightarrow \phi X ) 
= K \, \hat{\sigma}_{\rm LO} 
( gg \rightarrow \phi ) \int_{\tau}^1 {\tau \over x}~g(x,Q^2) g(\tau /x,
Q^2) ~dx ,
\end{equation}
where $\tau \equiv m_{\phi}^2 / s$ and $\sqrt{s}$ is the CM energy of the 
hadron colliders ($\sqrt{s} = 2$ TeV and 14 TeV for the Tevatron and 
LHC, respectively). The $K$ factor includes the QCD corrections, and we set
$K=1.5$.
The parton-level cross section for $gg\rightarrow \phi$ is given by 
\begin{equation}
\hat{\sigma}_{\rm LO} ( gg \rightarrow \phi ) = 
{\alpha_s^2 (Q) \over 256 \pi \Lambda_{\phi}^2} ~\left| b_{QCD} + 
\sum_q I_q ( x_q ) \right|^2, 
\end{equation}
where 
$I ( z )$ is given in the Eq.~(\ref{ffnt}). For the gluon distribution 
function, we use the CTEQ5L parton distribution functions \cite{cteq5}.
\begin{figure}
\centerline{\epsfxsize=7.5cm \epsfbox{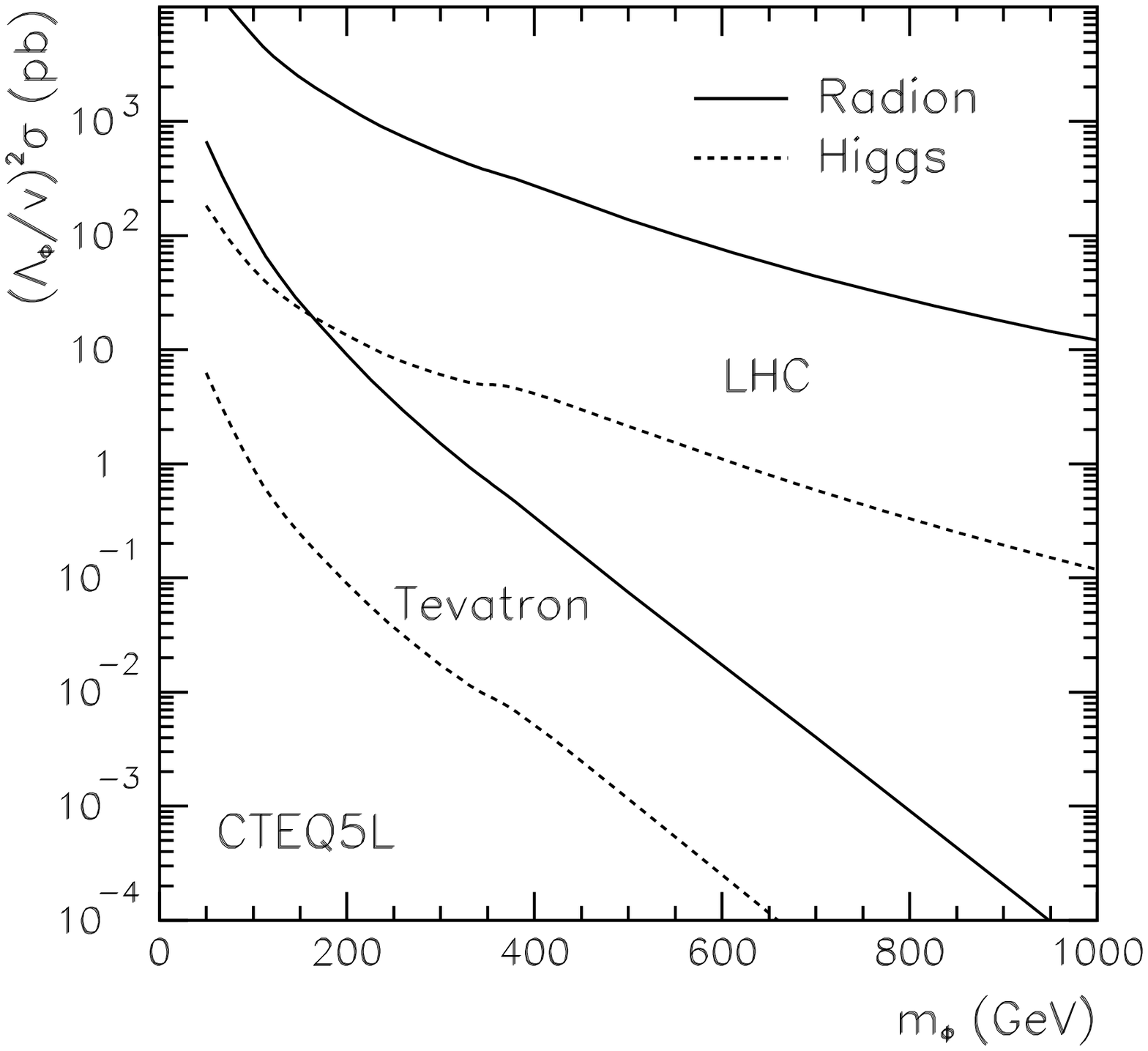}}
\caption{
The radion production cross section via gluon fusions at the Tevatron 
($\sqrt{s} = 2 $ TeV) and LHC ($\sqrt{s} = 14 $ TeV) with a scale factor 
$( \Lambda_{\phi} / v)^2$. The Higgs production cross sections are shown 
in dashed  curves for comparison. }
\label{hadron}
\end{figure}
In Fig.~\ref{hadron}, we show the radion production cross sections at the 
Tevatron and LHC as functions of $m_{\phi}$ for $\Lambda_{\phi} = v$. 
We set the renormalization scale $Q = m_{\phi}$ as shown in the figure. 
When we vary the scale $Q$ between $m_{\phi}/2$ and $2 m_{\phi}$, the 
production cross section changes about $+30 \%$ to $- 20 \%$. 
The production cross section will scale as $(v/\Lambda_{\phi})^2$ as before.
Compared to the SM Higgs boson productions, one can clearly observe that 
the trace anomaly can enhance the hadroproductions of a radion enormously. 
As in the SM Higgs boson, there is a great possibility to observe the radion 
up to mass $\sim 1 $ TeV if $\Lambda_{\phi} \sim v$.  For a smaller 
$\Lambda_{\phi}$, the  cross section becomes larger but the radion becomes 
very broader and it becomes more difficult to find such a scalar. For a larger
$\Lambda_{\phi}$, the situation becomes reversed : the smaller production
cross section, but a narrower width resonance, which is easier to detect.
In any case, however, one has to keep in mind that the perturbative unitarity
may be violated in the low $\Lambda_{\phi}$ region. 

At the $e^+ e^-$ colliders, the main production mechanism for the radion 
$\phi$ is the same as the SM Higgs boson : the radion--strahlung from $Z^0$ 
and the $WW$ fusion, the latter of which becomes dominant for a larger CM 
energy \cite{ee}. Again we neglect the anomaly contributions here.
Since both of these processes are given by the rescaling of 
the SM Higgs production rates, we can use the current search limits  
on Higgs boson to get the bounds on the radion.  
With the data from L3 collaboration \cite{l3}, we show the constraints of 
$\Lambda_\phi$ and $m_\phi$ in the left three curves of Fig.~\ref{bound}.
Since L3 data is for $\sqrt{s}=189$ GeV and the mass of $Z$ boson is 
about 91 GeV, 
the allowed energy for a scalar particle is about 98 GeV. If the mass of the
scalar particle is larger than 98 GeV, then the cross section vanishes. 
Therefore, if $m_\phi$ is larger than 98 GeV, there is no constraint on 
$\Lambda_\phi$. And the forbidden region in $m_\phi-\Lambda_\phi$ plane
is not changed by $m_h \gtrsim 98$ GeV, because there is no Higgs 
contribution to the constraint for $m_h \gtrsim 98$ GeV. 
\begin{figure}
\centerline{\epsfxsize=7.5cm \epsfbox{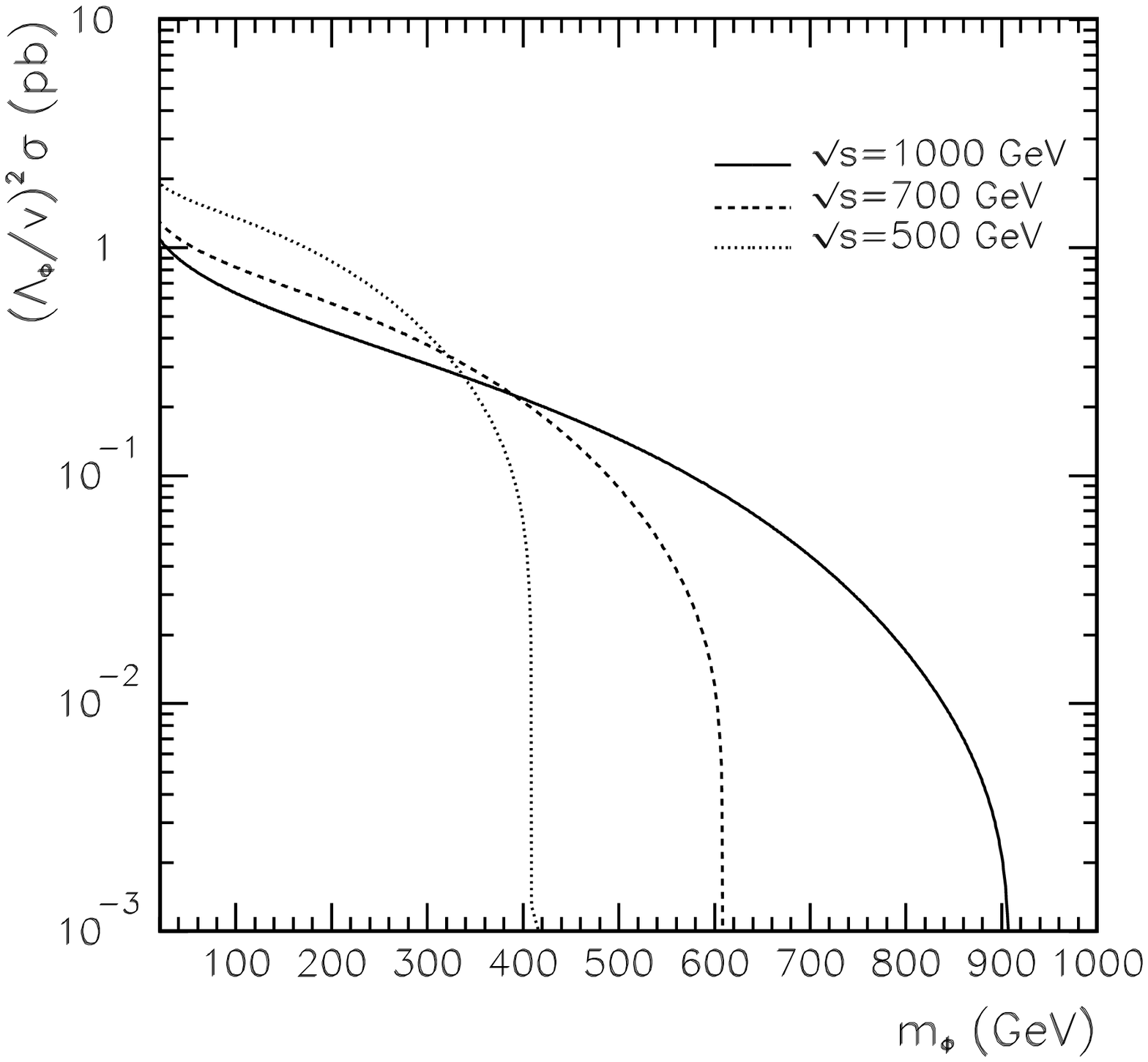}}
\caption{
The production cross section for the radion at NLC's at  
$\sqrt{s}$ = 500 , 700 and 1000 GeV, respectively.}
\label{lin}
\centerline{\epsfxsize=7.5cm \epsfbox{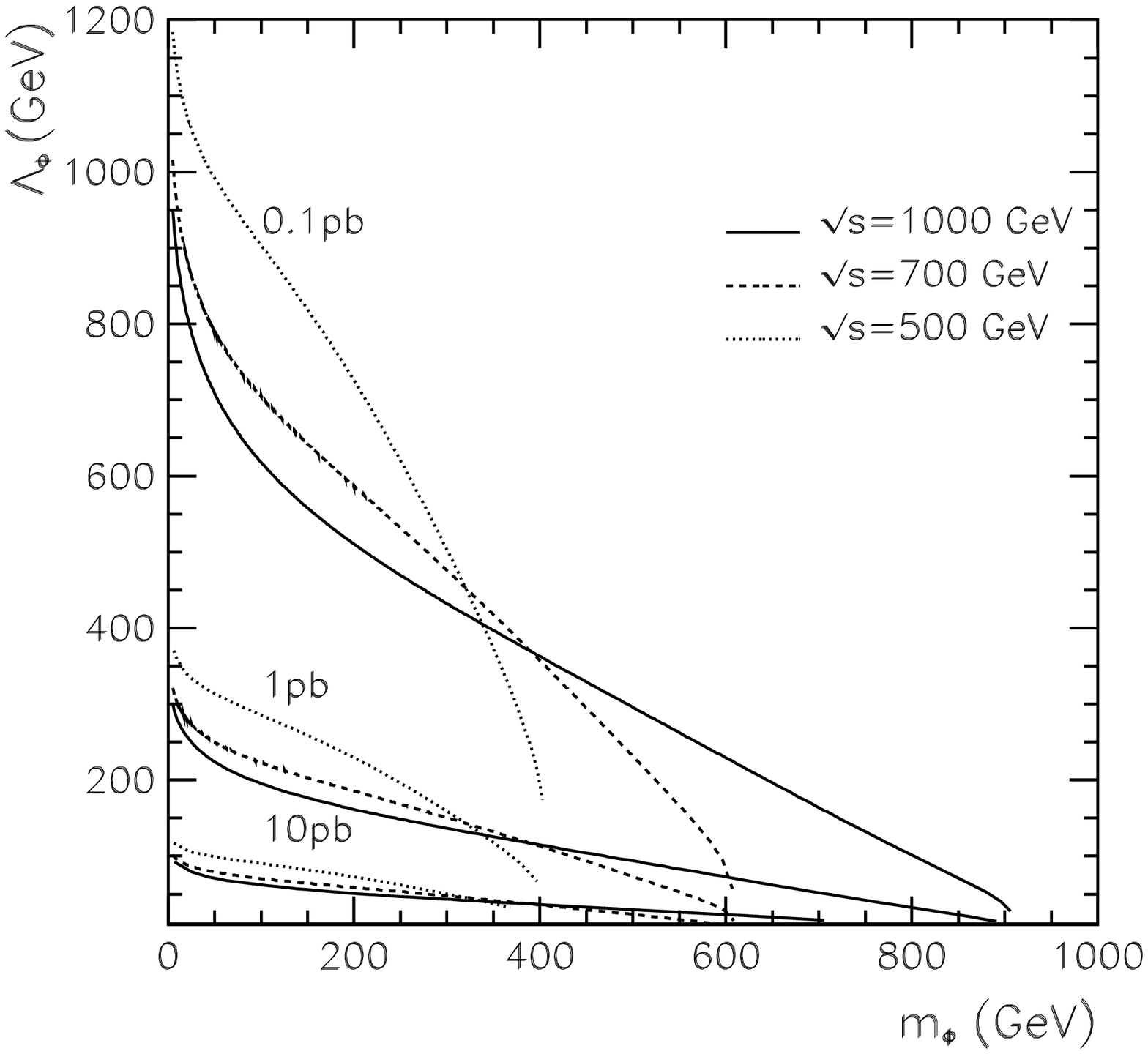}}
\caption{
The constant production cross section curves at next linear colliders
(NLC's)  for $\sqrt{s}$ = 500 , 700 and 1000 GeV }
\label{const-cs}
\end{figure}
The radion production cross sections at NLC's and the corresponding constant 
production cross section curves in the ($\Lambda_\phi$, $m_\phi$) plane are 
shown in Fig.~\ref{lin} and Fig.~\ref{const-cs}, respectively. We have chosen
three different CM energies for NLC's : $\sqrt{s} = 500$ GeV, 700 GeV and 1 
TeV. We observe that the relatively light radion ($m_{\phi} \lesssim 500$ GeV)
with $\Lambda_{\phi} \sim v$ (up to $\sim 1$ TeV) could be probed at NLC's 
if one can achieve high enough luminosity, since the production cross section 
in this region is less than a picobarn. 


There are also studies of the radion effects on low energy phenomenology 
such as muon $(g-2)$ \cite{mahanta} and the weak mixing angle \cite{jkim}. 
The effects are generally small in the region where perturbative unitarity 
is not violated. 

\section{Conclusions}

In summary, we demonstrated that the radion that stabilizes the fifth 
dimensional modulus in the RS I scenario has characteristic signatures at 
colliders due to the scale anomaly. Were it not for the scale anomaly in QFT, 
the radion would have behaved exactly the same as the SM Higgs except that 
the dimensionful parameter $v$ relevant for the SM Higgs is replaced by 
the radion decay constant $\Lambda_{\phi}$. The radion would have decayed
preferentially into $b \bar{b}$ final state for $m_{\phi} < 2 m_W$, and
into $WW/ZZ$ for a heavier radion ($m_{\phi} > 2 m_{V=W,Z}$), like the SM
Higgs boson. Because of the scale anomaly, however, the situation drastically
changes and $\phi \rightarrow g g $ is greatly enhanced over 
$\phi \rightarrow b \bar{b}$, and could be a dominant decay channel for a 
light radion. Also this enhanced $\phi-g-g$ ($\phi-\gamma-\gamma$) coupling 
makes the gluon (photon) fusion into a radion the dominant radion production
mechanism at hadron (photon) colliders.

\section*{Acknowledgments}
This work was supported in part by BK21 program (HSL), 
DFG-KOSEF exchange program (PK) and by KOSEF SRC program through CHEP at 
Kyungpook National University (PK).

\end{document}